\documentclass[twocolumn,showpacs,pas,prl,english]{revtex4}
\usepackage{amssymb}
\usepackage[T1]{fontenc}
\usepackage[latin1]{inputenc}
\usepackage{graphicx}

\makeatletter



\usepackage{babel}
\makeatother
\begin{document}

\title[]{Radiation reaction effects on ion acceleration in laser foil interaction}
\author{Min Chen, Alexander Pukhov, Tong-Pu Yu}
\address{Institut f\"{u}r Theoretische Physik I,
Heinrich-Heine-Universit\"{a}t D\"{u}sseldorf, 40225 D\"{u}sseldorf,
Germany}
%\ead{mchen911@ustc.edu}
\author{Zheng-Ming Sheng}
\address{Department of Physics, Shanghai Jiao Tong University,
Shanghai 200240, China \\and Beijing National Laboratory for
Condensed Matter Physics, Institute of Physics, Beijing 100080,
China}


\begin{abstract}
Radiation reaction effects on ion acceleration in laser foil
interaction are investigated via analytical modeling and
multi-dimensional particle-in-cell simulations. We find the
radiation effects are important in the area where some electrons
move backwards due to static charge separation field at the laser
intensity of $10^{22}~W/cm^2$. Radiation reaction tends to impede
these backwards motion. In the optical transparency region ion
acceleration is enhanced when the radiation effects are considered.
\end{abstract}

%Uncomment for PACS numbers title message
%\pacs{00.00, 20.00, 42.10}
% Keywords required only for MST, PB, PMB, PM, JOA, JOB?
%\vspace{2pc}
%\noindent{\it Keywords}: Article preparation, IOP journals
% Uncomment for Submitted to journal title message
%\submitto{\JPA}
% Comment out if separate title page not required
\maketitle

%\tableofcontents
%\newpage
\section{Introduction}
Along with the fast development of laser technology, the power
intensity which is inaccessible one decade before becomes reality.
The focused laser intensity of $10^{22}~W/cm^2$ is available today.
The new generation of laser system such as ELI is under
construction. Focused intensity of $10^{25}~W/cm^2$ is just
forward~\cite{ELI}. By use of such intense laser power, plenty of
physical problems can be studied and amount of applications are
waiting for discovering. Among them particle acceleration presently
and will still attract extensive attentions. Electron acceleration
by laser plasma accelerator is aimed to TeV~\cite{proton-driven},
ion acceleration also towards to stable, high and mono-energetic
beam~\cite{Macchi2009}. Besides ultrahigh power laser system, plasma
mirror technology also makes the high contrast clean pulse
possible~\cite{plasma-mirror}, which makes ultrathin
target~($\sim$nm thickness) now applicable for ion
acceleration~\cite{nm-target}. Unlike the target normal sheath field
acceleration~(TNSA)~\cite{ion-acce}, radiation pressure dominated
ion acceleration~(RPA) can have a much longer acceleration
distance~\cite{Robinson2008}. Simulation results show ions can get
GeV energy when a 20fs long laser pulse with $10^{22}~W/cm^2$
intensity is used~\cite{Chen2009}. Besides the two mentioned ion
acceleration mechanisms, recently ion acceleration in the optical
transparency region named the "laser breakout
afterburner"~(BOA)~\cite{Lin2007} has also been studied. It
demonstrates that ions can get continuously acceleration even the
laser pulse transmits through the ultrathin foil, which also gives a
possible way to get high energy ion beams.

As is well known, high energy particles when suffering acceleration
self-radiation is concomitant. Radiation reaction effects should be
considered when the radiation damping force is comparable to the
external one. For the above mentioned processes for particle
acceleration, electrons are always endured intensive acceleration.
When the laser intensity increases further, the radiation reactions
cannot be neglected. Naumova \textit{et al.} have studied the
radiation reaction effect on the laser hole boring process and point
out it plays a positive role as it allows one to maintain the
electron thermal energy on a relatively low level and prevents the
electron backward motion through the pulse~\cite{Naumova2009}.

In the present paper, we study their effects on ion acceleration. As
a primary result, we find the radiation mainly comes from the
electrons moving backwards in the laser pulse and the radiation
damping impedes this kind of backward motion, which can reduce the
particles' volume in the phase space and improve the ion
acceleration energy and quality.
\begin{figure}\centering
\begin{centering}\includegraphics[clip,width=60mm]{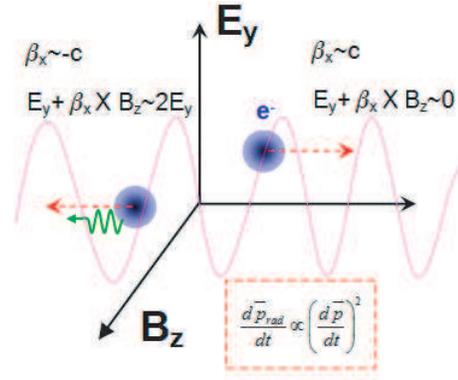} \par\end{centering}
\caption{(color online)\label{profile-fig1} Sketch map of laser
electron interaction. Here a linearly polarized laser pulse is
used.}
\end{figure}

\section{Radiation and its effect}
Before showing the effect of radiation damping, we first check the
threshold of the laser intensity for the important radiation
reaction effects from existed fundamental formulae. For an electron
with velocity $v{\sim}c$, one can get the radiation power $P(t)$ at
the local radiation time $t$ as:
\begin{equation}\label{radiation_parallel}
    P(t)=\frac{2e^2}{3m^2c^3}(\frac{d\vec{p}}{dt})^2,
\end{equation}
when $\vec{a}_{acce}{\parallel}\vec{v}$, and
\begin{equation}\label{radiation_parallel}
    P(t)=\frac{2e^2}{3m^2c^3}(\frac{d\vec{p}}{dt})^2\gamma^2,
\end{equation}
when $\vec{a}_{acce}{\perp}\vec{v}$. Here $\vec{p}$ is the electron
momentum, $\gamma$ is the relativistic factor, $e$ and $m$ are the
charge and mass of the electron, respectively. So the radiation due
to the transverse acceleration is $\gamma^2$ stronger than the one
due to the longitudinal acceleration. For the synchrotron radiation,
the radiation is mainly in the direction of electron motion and
concentrated within an angle of $\theta$ with
$\bigtriangleup\theta\sim\frac{1}{\gamma}$.

Based on the above knowledge, in the PIC code we only consider the
radiation due to the transverse acceleration and only the electrons
whose $\gamma\geq5$ are assumed to contribute the radiation.

By use of the normalized variables as in the PIC code ($p{\sim}p/mc,
t{\sim}t/T_0=\omega_0t/2\pi$), we can get the momentum variation
rate as:
\begin{equation}\label{momentum1}
   \frac{dp_{rad}}{dt}=\frac{2e^2}{3mc^2\lambda_0}(\frac{dp}{dt})^2\gamma^2.
\end{equation}
In one simulation step it changes:
\begin{eqnarray*}\label{momentum1}
% \nonumber to remove numbering (before each equation)
 (\frac{dp_{rad}}{dt})dt &=& (\frac{4}{9}\frac{e^2}{hc/2\pi}dp){\times}(\frac{3\gamma^2}{4\pi}\frac{dp}{dt}){\times}\frac{h\omega_0/2\pi}{mc^2}\\
   &=&
   n_{ph}\frac{\omega_c}{\omega_0}\frac{E_{ph0}(eV)}{5.11\times10^5},
\end{eqnarray*}
where $h$ is the Planck constant,  $\omega_0$ is the laser frequency
and $E_{ph0}$ is the photon energy of the laser pulse. To be simply,
in the PIC code we also assume the radiation is in form of photons.
The photon frequency is $\omega_c$ and the corresponding photon
number is $n_{ph}=8{\pi}e^2dp/9hc$. The radiation reaction force on
the electron then is:
\begin{eqnarray*}\label{momentum1}
% \nonumber to remove numbering (before each equation)
 \frac{dp_{rad}}{dt} &=& \frac{4}{9}\alpha{\times}\frac{3\gamma^2}{4\pi}{\times}\frac{E_{ph0}(eV)}{5.11\times10^5}(\frac{dp}{dt})^2\\
   &=&
   1.8791\times10^{-9}\gamma^2(\frac{dp}{dt})^2/\lambda_0({\mu}m).
\end{eqnarray*}
The reaction force points to the opposite direction of the electron
motion. On the other hand, the electron also feels the external
force as~(Here we only show the transverse force due to laser
field.):
\begin{equation}\label{momentum1}
   \frac{dp_{Laser\perp}}{dt}=2{\pi}q_e(\vec{E}+\vec{\beta}{\times}\vec{B})_\perp=-2\pi(\vec{E}_\perp+\vec{\beta_x}{\times}\vec{B}_\perp),
\end{equation}
where $E$ and $B$ are the intensities of electric and magnetic
fields normalized by $m{\omega_0}c/e$. We can get the threshold of
the laser intensity for the obvious radiation damping effect by
using:
\begin{equation}\label{momentum1}
   \frac{dp_{rad}}{dt}\sim\frac{dp_{Laser\perp}}{dt}.
\end{equation}
It is:
\begin{equation}\label{threshold}
   1.181\times10^{-8}\gamma^2(\vec{E}+\vec{\beta_x}{\times}\vec{B})/\lambda_0({\mu}m){\sim}1.
\end{equation}
Usually in a laser pulse, $|\vec{E}|=|\vec{B}|=a$. For an electron
moving along with a linearly polarized laser pulse, we have:
\begin{eqnarray*}
% \nonumber to remove numbering (before each equation)
  p_y &=& a \\
  p_x &=& a^2/2 > 0 \\
  \gamma &=& 1+a^2/2 \\
\end{eqnarray*}
Here $a$ is the normalized laser field. To get an obvious radiation
effect, it should satisfy ${\gamma}a{\geq}8.47\times10^7$,
correspondingly the laser intensity should satisfy: $a{\geq}550$. It
is about $4.2{\times}10^{23}~W/cm^2$ for a laser with a wavelength
of $1~{\mu}m$, which is higher than the current running laser
system. However, for an electron with longitudinal velocity of
$\beta_x\simeq0$, Eq.~(\ref{threshold}) changes to
$\gamma^2a\geq8.47\times10^7$. When the electron moves in the
opposite direction of the laser pulse with the longitudinal velocity
$\beta_x{\simeq}-1$, then one can get $\gamma^2a\geq4.23\times10^7$.
That is the threshold of $a$ can be $\gamma$ times smaller when the
electrons with the same energy~($\gamma$) move backward~(see
Fig.~\ref{profile-fig1}). The reason is that the radiation force is
proportion to the square of the acceleration force. For the
electrons moving forwards the transverse force they feel is about
zero~($\vec{E}+\vec{\beta_x}{\times}\vec{B}\approx0$), so the
radiation force is even smaller. However, for the electrons moving
backwards, the felt transverse acceleration force is about
$4{\pi}q_e\vec{E}$, so the radiation force is much larger.

\section{Simulation results and discussion}
\begin{figure}\centering
\begin{centering}\includegraphics[clip,width=85mm]{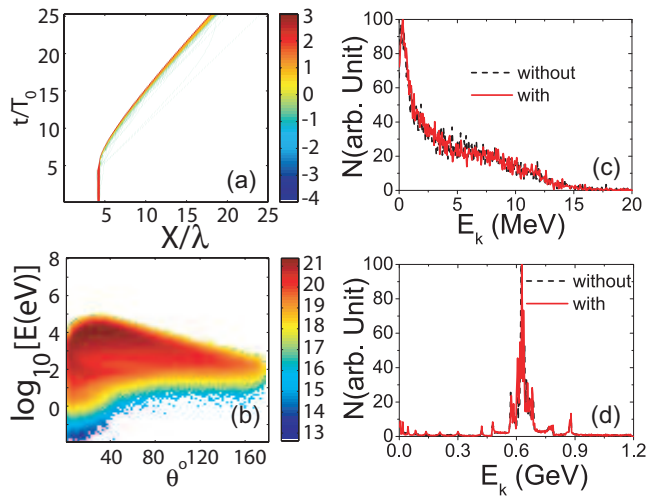} \par\end{centering}
\caption{(color online)\label{a100n100} (a) Spatial-temporal
distribution of the electron density. (b) Electron energy spectra in
the simulations with and without radiation damping effects at
$t=20~T_0$. (c) The energy angular distribution of the radiated
photons during $t=20~T_0$ and $t=21~T_0$. The colorbar shows the
relative photon number in the logarithmic value. (d) Proton spectra
in the simulations with and without radiation damping effects at
$t=20~T_0$. Here $a=100$ and $n=100$.}
\end{figure}
In the following, we use PIC simulations to check the radiation
reaction effects and show their importance on ion acceleration. To
simulate the damping effects we suppose that, at any given moment of
time, the electron radiation spectrum is synchrotron
like~\cite{kiselev-synchrotron}. The critical frequency $\omega_c$
is given by the relation
$\omega_c=(3/4\pi)\gamma^2|{\bigtriangleup}P_\perp|/(dt)$;
${\bigtriangleup}P_\perp$ is the variation of transverse electron
momentum force during the time step of $dt$. In our PIC code, we
follow trajectories of each electron and calculate the emission
during the interaction. We calculate the damping effects by
considering the electron's recoil due to the emitted radiation. The
recoil force is included in the equations of electron motion. We
should mention that our method is different with the one used by
Martins \textit{et al.} in OSIRIS code, in which they can get the
radiation field and frequency in a faraway virtual
detector~\cite{Martins2009}. In our code we do not pay attention to
the received radiation field on the virtual detector. Our main
attention is focused on the reaction effects on plasma itself. The
radiation we recorded every simulation step is the photons radiated
in the local time of radiation, not the one at the observation time.
Our method is also simpler than the one used by Sokolov \textit{et
al.}, in which the modified non-perturbative Lorentz-Abraham-Dirac
equation is resolved for particle motion instead of the normal
Lorentz equation~\cite{Sokolov2009}. Although with our method the
code cannot gives a correct radiation field on a virtual detector,
it is relatively simple and can give appropriate description for the
plasmas under radiation damping. Our findings are similar with
Naumova \textit{et al.} as they study the hole boring process for
fast ignition, however, our main interest is on the ion acceleration
in the laser foil interaction.

We take 1D-PIC simulations with the KLAP-1D code~\cite{KLAP}. In the
simulation, the laser pulse has trapezoidal temporal intensity
profile ("linear increase-plateau-linear decrease") with the pulse
length $1\lambda/c$-$16\lambda/c$-$1\lambda/c$. Here
$\lambda=1~{\mu}m$ is the laser wavelength. The normalized laser
electric field $a=eE/m{\omega_0}c$ changing from 100 to 500 are
used. The target plasma are composed of electrons and protons and
they are initially uniformly distributed from $x=4~\lambda$ to
$x=4.3~\lambda$. The plasma density is fixed to be $100~n_c$ with
the critical density $n_c=1.1{\times}10^{21}/cm^3$.
\begin{figure}\centering
\begin{centering}\includegraphics[clip,width=85mm]{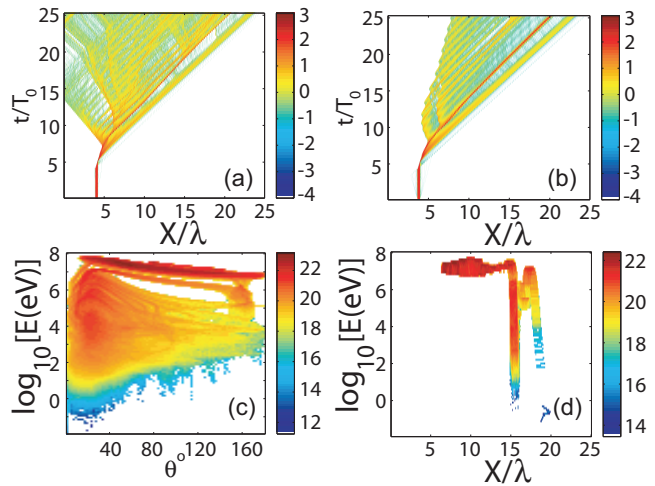} \par\end{centering}
\caption{(color online)\label{ne_and_spec} (a) Spatial-temporal
distribution of the electron density for the simulation without
radiation damping effects. (b) The case with radiation damping
effect. (c) The energy angular distribution of the radiated photons
during $t=20~T_0$ and $t=21~T_0$. The colorbar shows the relative
photon number in the logarithmic value. (d) The radiated photon
energy and positions during $t=20~T_0$ and $t=21~T_0$. The colorbar
shows the relative photon number in the logarithmic value. Here
$a=180$ and $n=100$.}
\end{figure}

For the presently fixed plasma density we find when the laser
intensity is lower than $a{\sim}125$ the radiation effects on the
particle energy spectrum is not observable. Fig.~\ref{a100n100}(a)
shows the spatial temporal distribution of the electron density in
the simulation of $a=100$. In this condition, the ions are
accelerated in the radiation pressure dominated region. Electrons
and ions are moving together and the acceleration is phase stable.
No obvious electron backward motion happens. The electron energy
spectrum does not vary so much between the two simulations with and
without radiation damping as shown in Fig.~\ref{a100n100}(c). There
is a sharp peak in the proton spectrum with the position at 0.63~GeV
after $16~T_0$ acceleration time which is close to the theoretical
value of 0.66~GeV. Fig.~\ref{a100n100}(c) shows the energy angular
distribution of the radiated photons. Most of the radiation is in
the forward direction and the photon energies are about 10~KeV. From
the simulation we calculate the average radiation power is about
$1.64\times10^{17}~W/cm^2$ which is negligible compared to the laser
power of $2.76\times10^{22}~W/cm^2$.

When we increase the laser intensity further to $a=180$, the
radiation damping effects obviously appears.
Fig.~\ref{ne_and_spec}(a,b) show the spatial temporal distribution
of the electron density in the two compared simulations~(with and
without radiation effects). As we see when there is no radiation
effects amount of electrons move backwards and the electron density
distribution disperses in space. On the contrary, after considering
the radiation effects this kind of backward acceleration has been
suppressed. More electrons are concentrated and move along with the
ion bunch as shown in Fig.~\ref{ne_and_spec}(b). We show the energy
angular distribution of the radiated  photons in
Fig.~\ref{ne_and_spec}(c). It composes of two parts. Except the
lower part with low photon energy which is similar with the one in
Fig.~\ref{a100n100}(c) (We call this "low energy radiation".), there
is a high energy part whose radiation direction is uniformly
distributed in the forward and backward directions (We call this
"high energy radiation"). To find the source electrons for these
radiated photons, we show the radiation position distribution in
Fig.~\ref{ne_and_spec}(d). Comparing with the positions of the
electrons at $t=20~T_0$ in Fig.~\ref{ne_and_spec}(b), we can see the
low energy radiation ($E_{photon}<1~MeV$) mainly comes from the
electrons moving with the ion bunch and the radiation mainly
concentrates within the angle of $20^o$, the high energy
photons~($E_{photon}{\sim}22.5~MeV$) are radiated by the electrons
which move behind the ion bunch where the plasma is transparent to
the laser pulse and electrons move both forwards and backwards, the
radiation is also almost uniformly in the angular distribution. The
total radiation power density is about $8.84\times10^{21}~W/cm^2$
which is almost $10\%$ of the laser power density.

\begin{figure}\centering
\begin{centering}\includegraphics[clip,width=85mm]{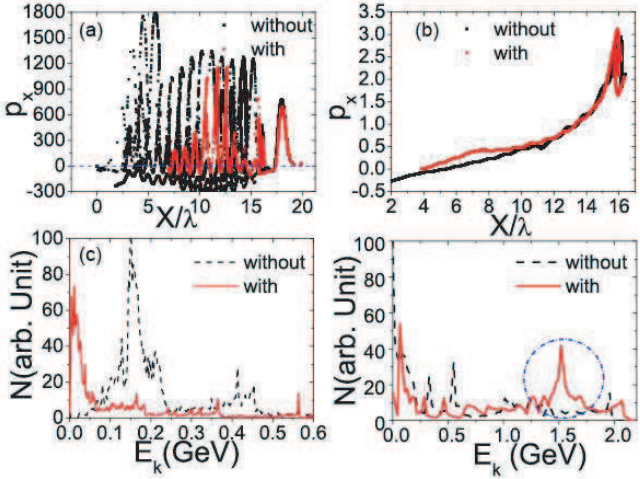} \par\end{centering}
\caption{(color online)\label{particle_spec} (a) Electron
distribution in the $x-p_x$ phase space, the momentum is normalized
by $m_ec$; (b) Proton distribution in the $x-p_x$ phase space; (c)
Electron energy spectrum, the momentum is normalized by $m_ic$; (d)
Proton energy spectrum. The time here is $t=20~T_0$.}
\end{figure}

Particle distributions in the phase space are shown in
Fig.~\ref{particle_spec}(a,b). Electrons have much smaller volume in
the phase space when the radiation damping is included in the
simulation. In this case no backward ion acceleration has been
found. Much more ions are confined in the front accelerating bunch
compared with the simulation without radiation damping. The electron
energy spectrum also shows that electrons have a much lower peak
energy (10~MeV) in the radiation included simulation, however, it is
about 150~MeV in the simulation without radiation damping. Proton
energy spectrum shown in Fig.~\ref{particle_spec} confirms much more
protons are concentrated and accelerated in the bunch whose peak
energy is about 1.5GeV at $t=20~T_0$.

From the energy point of view, at $t=20~T_0$, totally
$5.01\times10^9~J/cm^2$ laser energy density has been transported
into the simulation box. When no radiation damping is included,
among them $2.44\%$ ($1.22\times10^8~J/cm^2$) transforms to
electrons and $6.93\%$ ($3.47\times10^8~J/cm^2$) transforms to
protons. When radiation damping is included, $1.23\%$
($6.15\times10^7~J/cm^2$) transforms to electrons and $11.16\%$
($5.59\times10^8~J/cm^2$) transforms to protons. As we see radiation
damping reduces the electron energy, however, it improves the proton
acceleration. Neglect of radiation damping gives a lower estimation
of energy conversion efficiency and worse spectrum.

In our simulations, we find that the high energy radiation appears
when $a>125$. Its quotient among the total radiation increases with
the laser intensity. When $a>300$, there is only high energy
radiation. These different radiation profiles also reflect the
different acceleration scenarios. From the balance between the
forces due to electrostatic field and radiation
pressure~\cite{Chen2009}:
\begin{eqnarray}
% \nonumber to remove numbering (before each equation)
 {\pi}n_0^2l^2 &=& \frac{a_0^2}{\pi}\frac{1-\beta_e}{1+\beta_e},
\end{eqnarray}
where $l$ is the thickness of the target normalized by laser wave
length ($\lambda_0$) and $\beta_e=a/(a+\sqrt{m_in_0})$ is the
relativistic hole boring velocity~\cite{Robinson2009}, we can get
the critical laser intensity for the laser pulse transmitting
through the target. For our present simulation parameters: $n_0=100,
l=0.3, m_i=1836$, we get $a_{cr}~{\approx}~117$. The intensity is
close to the critical value ($a~{\approx}~125$) for the high energy
radiation. It means when the laser intensity is higher than
$a_{cr}$, the target tends to be transparent to the laser pulse.
Stable (at least in the 1D case) structure of radiation dominated
ion acceleration begins mixing with other acceleration mechanism
(such as BOA). However, some ions can still be accelerated by the
radiation pressure. The amount of these ions depends on the laser
intensity. Other ions are accelerated or decelerated by the
dispersed electrons. When the laser intensity increases further,
radiation pressure acceleration disappears completely. Ions can only
be accelerated in the heated electron pool which moves with the
laser pulse. In this scenario, the radiation damping is important
and necessary, which actually makes the electron pool cooling down
and improves the ion acceleration. Further investigation still needs
for this kind of acceleration scenario.

\section{Summary and discussion}
In summary, by use of PIC simulations we studied the radiation
reaction effects on the ion acceleration in the radiation pressure
dominated region and the transparent plasma region. We find
radiation damping effects are important for the electrons moving
backwards and immersed in the laser pulse. Self-radiation impedes
backward motion, cools down the electrons and makes more ions be
concentrated and accelerated. We notice recently many studies on ion
acceleration by use of ultrathin foil target, in which target plasma
is almost transparent to the laser pulse. Electrons are dancing with
the transmitted relatively long laser pulse. The local charge
separation field takes in charge of ion acceleration. In this
condition, a correct electron distribution in phase space is
important to get the correct final maximum accelerated ion energy
and acceleration scaling. When the laser intensity is larger than
$a=100$, radiation reaction could change the electron distribution
in phase space. Furthermore, with future laser system such as
ELI~($a>1000$), laser intensity is high enough to awake the
contribution of radiation damping. Even for the electrons move
forward, the radiation reaction should be considered. It deserves
and is necessary to include the radiation reaction effects in the
future PIC simulations when ultra intense laser pulse is used.

\section{Acknowledgements}
This work is supported by the DFG programs TR18 and GRK1203. MC
acknowledges support by the Alexander von Humboldt Foundation. ZMS
is supported in part by the National Nature Science Foundation of
China (Grants No. 10674175, 10734130) and the National Basic
Research Program of China (Grant No. 2007CB815100).

\end{document}